\pdfoutput=1
\documentclass[11pt,a4paper]{article}

\usepackage[utf8]{inputenc}
\usepackage[T1]{fontenc}

\usepackage{amsmath,amssymb,amsfonts}
\usepackage{bm}
\usepackage{graphicx}
\usepackage{geometry}
\usepackage{placeins}

\usepackage{authblk}
\usepackage[numbers,sort&compress]{natbib}

\usepackage{doi}
\usepackage{hyperref}

\hypersetup{
    colorlinks=true,
    linkcolor=blue,
    citecolor=blue,
    urlcolor=blue,
    breaklinks=true
}

\newcommand{\DXICDELTA}{0.02}      
\newcommand{\DXICVAL}{\pi/2}       
\newcommand{\DXICROB}{6.6}         

\newcommand{\NGRIDVALS}{2.5, 3.0, 3.5, 4.0, 4.5}
\newcommand{\DGRIDVALS}{0.02, 0.04, 0.06, 0.08, 0.10}
\newcommand{\CHIRPVAL}{0.5}

\title{\textbf{Generalized Thermodynamics of Solitonic Event Horizons in Dispersive Field Theories}}

\author[1,2]{\textbf{Hasan Oguz}}
\affil[1]{\small Department of Computer Technologies, Vocational School, Istanbul Okan University, 34959 Tuzla, Istanbul, T\"urkiye}
\affil[2]{\small Material Physics Simulation Laboratory, Science Faculty, Pamukkale University, 20160 Pamukkale, Denizli, Türkiye}
\affil[ ]{\textit{Email: hasan.oguz@okan.edu.tr}}

\date{}

\begin{document}
\maketitle

\begin{abstract}
\noindent
The realization of Hawking radiation in optical analogs has historically focused on kinematic observables, such as the effective temperature determined by the horizon's surface gravity. A complete thermodynamic description, however, necessitates a rigorous definition of entropy and irreversibility, which has remained elusive in Hamiltonian optical systems. In this work, we bridge this gap by introducing an operational entropy for solitonic event horizons, derived from the spectral partitioning of the optical field into coherent solitonic and incoherent radiative subsystems.  The emission of resonant radiation, driven by the breaking of soliton integrability under higher-order dispersion, is the fundamental mechanism for entropy production.  Numerical simulations of the generalized nonlinear Schr\"odinger equation (GNLSE) demonstrate that, in a coarse-grained sense, this process obeys a generalized second law (GSL), $\Delta S_{\mathrm{tot}} \ge 0$, robustly across a wide range of soliton orders and dispersion strengths.  These results show that event horizons in dispersive field theories behave as consistent nonequilibrium thermodynamic systems, and that the relevant entropy is accessible from laboratory spectral measurements.
\end{abstract}

\vspace{1em}

\section{Introduction}
Analog gravity exploits the formal equivalence between linearized excitations in structured media and quantum fields propagating on curved spacetime backgrounds. Since Unruh's seminal observation that sound waves in an inviscid fluid obey a covariant wave equation determined by an effective metric \cite{Unruh1981}, a broad class of analog systems has emerged. While the underlying microscopic physics differs across platforms, these systems provide a unified laboratory for probing horizon kinematics and quantum emission processes, most notably Hawking radiation \cite{Hawking1974}, without recourse to astrophysical black holes, a domain where the generalized terminology and optical analogs have been broadly surveyed \cite{Aguero2020}.

Acoustic analogs currently represent the most experimentally mature realization. In Bose--Einstein condensates (BECs), phononic excitations in the hydrodynamic limit of the Gross--Pitaevskii equation experience a curved metric, while Bogoliubov dispersion provides a natural ultraviolet regularization \cite{Garay2000}. Recent experiments have reported thermal phonon spectra and nonlocal density--density correlations consistent with spontaneous Hawking pair production \cite{Steinhauer2016}. Furthermore, pioneering theoretical frameworks for quantum backreaction \cite{Baak2022} and cosmological particle production \cite{Fischer2004_Inflation, Fischer2004_Cosmo} in ultracold gases have provided a rigorous basis for understanding the dynamical interplay between horizon geometry and quantum fluctuations. These results confirm the quantum-thermodynamic nature of analog horizons, yet acoustic platforms remain largely restricted to scalar excitations, though recent proposals have extended these systems to 2+1D geometries to probe Penrose-type singularities \cite{Datta2025_Penrose}.

Photonic analog gravity, by contrast, exploits dispersion engineering in dielectric media. In nonlinear optical fibers, an ultrashort soliton induces a co-moving refractive-index perturbation via the Kerr effect, defining an effective spacetime for weak probe fields \cite{Leonhardt2000,Philbin2008}. An optical horizon forms when the probe group velocity matches the soliton velocity, leading to mode conversion and Hawking-like radiation \cite{Belgiorno2010,Rubino2012,Webb2014}. Beyond fibers, photonic crystals and coupled-resonator structures allow for stationary horizons associated with frozen light and rainbow trapping. However, unlike acoustic systems which often approximate Lorentzian geometry, photonic platforms are intrinsically open, driven, and highly dispersive. This fundamental nonequilibrium nature complicates the definition of vacuum states and entropy.

Consequently, a significant conceptual gap remains: while the kinematics of optical horizons (surface gravity, Hawking temperature) are well established \cite{Unruh1995,Robertson2012}, their \emph{thermodynamic} characterization is incomplete. In gravitational physics, horizon temperature acquires meaning only through its connection to entropy and the generalized second law (GSL) \cite{Bekenstein1973,Bekenstein1974}. While fluid analogs have successfully probed these connections via phonon entanglement \cite{Steinhauer2016} and information scrambling \cite{Fischer2022_Scrambling}, a comparable thermodynamic framework for optical solitons, namely a precise definition of \emph{what} counts as entropy in a pulse-propagation experiment, has been notably absent.

In this work, we address this gap by formulating a generalized thermodynamic framework for solitonic optical event horizons. A direct transplantation of entropy concepts from fluid dynamics is insufficient; instead, we formulate entropy operationally through coarse-graining over inaccessible degrees of freedom \cite{Breuer2002}. In nonlinear optical fibers, third-order dispersion breaks the integrability of the nonlinear Schr\"odinger equation, coupling the soliton to a continuum of dispersive modes known as resonant radiation \cite{Akhmediev1995,Skryabin2010}. By identifying this resonant radiation as the primary entropy carrier, we demonstrate via numerical simulation that the system obeys the GSL.  This shows photonic analog gravity admits a consistent nonequilibrium thermodynamic description, not only a kinematic one.

From the standpoint of gravitational physics, the defining conceptual role of an event horizon is not associated with a particular microscopic realization, but with the emergence of causal separation and irreversible information flow across a null boundary. This viewpoint is formalized in black hole thermodynamics through the GSL, which asserts that the sum of horizon entropy and exterior entropy is non-decreasing under physically admissible processes \cite{Bekenstein1974,Bekenstein1981,Wall2010}. While the geometric area law linking entropy to horizon area is specific to Einstein gravity \cite{Bekenstein1973}, the logical content of the GSL is more general, relying on coarse-graining over inaccessible degrees of freedom and on the existence of a horizon that mediates information exchange between subsystems. This structural interpretation has been emphasized in studies of horizon entropy as entanglement entropy \cite{Bombelli1986,Srednicki1993} and underlies the expectation that GSL behavior is not intrinsically tied to the microscopic details of quantum gravity.

The coarse-grained GSL established below rests on only three structural ingredients: a localized background acting as a horizon, coarse-graining over modes that become operationally inaccessible, and an irreversible spectral-redistribution channel. None of these is specific to solitons, so any integrability-broken dispersive horizon possessing the same ingredients is expected to exhibit analogous behavior; a quantitative demonstration of this broader universality lies beyond the integrability-breaking soliton-fission scope of the present analysis.

Within this broader perspective, the solitonic event horizon investigated here  is a dispersive, nonequilibrium horizon arising in an integrability-broken field theory. Such horizons are characterized by a localized background that induces mode conversion, a continuum of outgoing dispersive channels that effectively function as an environment, and a dynamical mechanism for irreversible spectral redistribution \cite{Unruh1995,Visser1998,Robertson2012}. In optical systems, these ingredients naturally emerge in the interaction of solitons with higher-order dispersion, leading to resonant radiation and broadband mode coupling \cite{Akhmediev1995,Dudley2006,Skryabin2010,Husko2013}. The entropy introduced in this work should therefore be interpreted as an operational, coarse-grained quantity rather than a microscopic state-counting entropy, analogous to entanglement- or environment-induced entropy in open systems \cite{Breuer2002}. The observed  coarse-grained increase of the total entropy does not depend on quantum statistics or gravitational dynamics, but on the universal  coupling between horizons, coarse-graining, and irreversibility. As a result, the conclusions are expected to apply generically to analog horizon systems, including Bose--Einstein condensates and hydrodynamic flows \cite{Garay2000,Weinfurtner2011,Steinhauer2016}, and to inform the interpretation of generalized second-law behavior beyond the specific optical realization considered.

\section{Thermodynamic Formalism}
\label{sec:formalism}
We adopt an operational approach to entropy, appropriate for driven, nonequilibrium systems in which a microscopic definition of thermodynamic entropy is neither accessible nor meaningful. Entropy is therefore defined through coarse-graining, quantifying the irreversible redistribution of spectral weight under a resolution-limited description \cite{Breuer2002}. Throughout this work, all entropic quantities are expressed in natural units (nats), corresponding to logarithms taken with base $e$.

Operationally, this construction is equivalent to tracing over dispersive modes outside a resolution-limited observational algebra, yielding an effective reduced description analogous to entanglement entropy in quantum field theory on curved spacetime \cite{Bombelli1986,Wall2010}. This open-system reading of the GSL is precisely the Prigogine-type nonequilibrium framework recently connected to Hawking radiation \cite{Sonnino2024a}.

The central idea is to partition the optical field into two dynamically coupled subsystems: a coherent, localized solitonic component and an incoherent radiative component generated through integrability-breaking dispersion. This partition defines an effective system--bath structure, enabling an operational formulation of entropy production. The soliton-dominated domain $\Omega_S$ and the complementary radiation domain $\Omega_R$ are disjoint, $\Omega_S\cap\Omega_R=\emptyset$, and together cover the active spectral support; the field and spectral variables are specified at first use in Eq.~\eqref{eq:pS} below.

\subsection{Effective Horizon Entropy}
The soliton constitutes a spectrally localized structure in the comoving time coordinate $\tau$, acting as an effective horizon for dispersive modes. 

To characterize the information content of this coherent subsystem on exactly the same footing as the radiative subsystem, we introduce the normalized spectral distribution over the soliton-dominated region $\Omega_S$. Here $\psi(\tau,\xi)$ is the slowly varying complex field envelope, with $\tau$ the comoving (retarded) time and $\xi$ the normalized propagation coordinate; $\tilde{\psi}(k,\xi)=\mathcal{F}_\tau[\psi(\tau,\xi)]$ is its Fourier transform with respect to $\tau$, $k$ being the dimensionless spectral wavenumber; and $p_S(k,\xi)$, defined in Eq.~\eqref{eq:pS}, is the resulting normalized spectral probability density on $\Omega_S$:
\begin{equation}
\label{eq:pS}
p_S(k,\xi) =
\frac{|\tilde{\psi}(k,\xi)|^2}{\displaystyle \int_{\Omega_S} |\tilde{\psi}(k',\xi)|^2 \, dk'},
\qquad
\int_{\Omega_S} p_S(k,\xi)\, dk = 1 ,
\end{equation}
and define the effective horizon entropy as its Shannon (differential) entropy,
\begin{equation}
S_{\mathrm{hor}}(\xi)
=
- \int_{\Omega_S} p_S(k,\xi)\, \ln p_S(k,\xi)\, dk .
\end{equation}
This definition is structurally identical to the radiation entropy introduced below; $S_{\mathrm{hor}}$ and $S_{\mathrm{rad}}$ are therefore commensurate information-theoretic quantities, rather than a photon number and an entropy. The exponential
\begin{equation}
\mathcal{N}_S(\xi) = \exp\!\left[\,S_{\mathrm{hor}}(\xi)\,\right]
\end{equation}
is the spectral participation number of the soliton: it measures the effective number of occupied spectral cells of the coherent subsystem and plays the role of an information-theoretic horizon ``size,'' without invoking any microscopic state counting. As resonant radiation is emitted, the soliton spectrum reorganizes and broadens; $S_{\mathrm{hor}}$ therefore quantifies the spectral compactness of the horizon subsystem rather than its energy content, and is no longer constrained to decrease monotonically. Both entropies are defined up to an additive constant $\ln\Delta k$ fixed by the spectral resolution; this constant cancels identically in the entropy differences and in the total balance below, consistent with the operational character of the construction. This depletion-and-reorganization of the coherent subsystem is formally analogous to the dynamics of horizon area in semiclassical black hole thermodynamics \cite{Bekenstein1973}, though no assumption of equilibrium is invoked.

\subsection{Spectral Entropy of Radiation}
The emitted resonant radiation forms an effectively mixed subsystem due to phase decoherence induced by dispersion and nonlinear mode coupling. We define a normalized spectral probability density over the radiation-dominated region $\Omega_R$ as
\begin{equation}
p(k,\xi) =
\frac{|\tilde{\psi}(k,\xi)|^2}{\displaystyle \int_{\Omega_R} |\tilde{\psi}(k,\xi)|^2 \, dk},
\qquad
\int_{\Omega_R} p(k,\xi)\, dk = 1 .
\end{equation}
The operational spectral entropy of the radiation field is then given by the Shannon entropy
\begin{equation}
S_{\mathrm{rad}}(\xi)
=
- \int_{\Omega_R} p(k,\xi)\, \ln p(k,\xi)\, dk .
\end{equation}
This entropy quantifies the degree of spectral delocalization induced by irreversible information transfer from the soliton to the dispersive bath. As in Page's analysis of black hole entropy \cite{Page1993}, the growth of $S_{\mathrm{rad}}$ reflects information redistribution under subsystem tracing rather than thermal equilibration. Irreversibility in the present framework does not arise from dissipation or coupling to an external reservoir; instead, it emerges from dynamical coarse-graining induced by integrability breaking and subsequent subsystem tracing. The higher-order dispersion term couples the initially integrable soliton dynamics to a continuum of dispersive modes, generating correlations that are operationally inaccessible under a reduced description. Entropy production therefore reflects information loss under coarse-grained observation, rather than any microscopic violation of unitarity.

\subsection{Generalized Second Law}
\label{subsec:gsl_def}
The total coarse-grained entropy of the optical horizon system is defined as the sum of the solitonic and radiative contributions,
\begin{equation}
\label{eq:stot_def}
S_{\mathrm{tot}}(\xi)
=
S_{\mathrm{hor}}(\xi)
+
S_{\mathrm{rad}}(\xi) .
\end{equation}

Our central hypothesis is that the irreversible emission of resonant radiation enforces the GSL in a coarse-grained sense. The instantaneous rate $dS_{\mathrm{tot}}/d\xi$ is \emph{not} sign-definite: over a single breathing cycle of the (fissioning) higher-order soliton, the spectrum of the coherent subsystem reversibly contracts and expands, and these reversible excursions carry no net entropy. (Breathing and fission are defined quantitatively in Sec.~\ref{subsec:fission}.) The physically meaningful statement therefore concerns averages over a coarse-graining scale $\Delta\xi_c$ set by the internal fission dynamics, whose operational determination from the simulation data is also given in Sec.~\ref{subsec:fission}: the coarse-grained entropy is non-decreasing,
\begin{equation}
\label{eq:gsl_coarse}
S_{\mathrm{tot}}(\xi_2) \ge S_{\mathrm{tot}}(\xi_1)
\quad \text{for} \quad
\xi_2 - \xi_1 \gtrsim \Delta\xi_c ,
\qquad
\left\langle \frac{d S_{\mathrm{tot}}}{d\xi} \right\rangle_{\Delta\xi_c} \ge 0 .
\end{equation}
To make this quantitative we introduce the entropy-production efficiency
\begin{equation}
\label{eq:eta}
\eta_{\mathrm{GSL}}(\xi)
=
\frac{\displaystyle \int_0^{\xi} \max\!\left(0,\, dS_{\mathrm{tot}}/d\xi'\right)\, d\xi'}
     {\displaystyle \int_0^{\xi} \left| dS_{\mathrm{tot}}/d\xi' \right|\, d\xi'} ,
\end{equation}
which equals $1$ for strictly monotonic growth and $\tfrac{1}{2}$ for an unbiased random walk. A coarse-grained GSL holds whenever two conditions are met simultaneously: (i)~a strictly positive net production over the full propagation, $\Delta S_{\mathrm{tot}}=S_{\mathrm{tot}}(\xi_f)-S_{\mathrm{tot}}(0)>0$, and (ii)~$\eta_{\mathrm{GSL}}>\tfrac{1}{2}$, i.e., entropy-increasing steps dominate entropy-decreasing ones along the trajectory. Verification of conditions (i)--(ii) constitutes the thermodynamic criterion by which solitonic optical event horizons are shown to function as genuine nonequilibrium analogs of gravitational horizons, despite the Hamiltonian nature of the underlying field equations.

\section{Numerical Framework}
\label{sec:numerics}
Pulse propagation is governed by the dimensionless generalized nonlinear Schr\"odinger equation (GNLSE)  \cite{Agrawal2013,Younas2023},

\begin{equation}
\label{eq:dimensionless_gnlse}
\partial_\xi \psi
=
i \left(
\frac{1}{2}\partial_\tau^2
+ |\psi|^2
+ i \delta_3 \partial_\tau^3
\right)\psi ,
\end{equation}

where $\xi$ denotes the normalized propagation coordinate, $\tau$ is the comoving time variable, and $\delta_3$ represents the normalized third-order dispersion coefficient responsible for integrability breaking and resonant radiation emission. Equation~\eqref{eq:dimensionless_gnlse} is written in the anomalous-dispersion convention supporting a stable bright soliton, with the third-order-dispersion term carrying the sign $+i\delta_3\partial_\tau^3$; linear waves $\psi\propto e^{ik\tau}$ then evolve under the spectral operator $D(k)=i\bigl(-\tfrac12 k^2+\delta_3 k^3\bigr)$, accumulating the propagation constant $K_{\mathrm{rad}}(k)=-\tfrac12 k^2+\delta_3 k^3$ per unit~$\xi$.

\emph{Cherenkov (resonant-radiation) phase matching} denotes the condition under which such a linear dispersive wave co-propagates in phase with the soliton, so that energy is resonantly and unidirectionally transferred from the soliton to that wave; it is the fiber-optic analog of Cherenkov emission by a moving source \cite{Wai1986,Akhmediev1995,Skryabin2010}. A fundamental soliton of amplitude $A$ accumulates the nonlinear propagation constant $K_{\mathrm{sol}}=A^2/2$, so the resonance occurs at the wavenumber $k_{\mathrm{RR}}$ solving
\begin{equation}
\label{eq:phasematch}
K_{\mathrm{rad}}(k_{\mathrm{RR}})=K_{\mathrm{sol}}
\quad\Longleftrightarrow\quad
\delta_3\, k_{\mathrm{RR}}^3-\tfrac12\, k_{\mathrm{RR}}^2=\frac{A^2}{2}.
\end{equation}
For $\delta_3\ll1$ the relevant root lies at $k_{\mathrm{RR}}=c_{\mathrm{RR}}/\delta_3$, with an $\mathcal{O}(1)$ coefficient $c_{\mathrm{RR}}$ fixed by the cubic--quadratic balance in Eq.~\eqref{eq:phasematch} and weakly corrected by the amplitude of the radiating (sub-)soliton; this $\delta_3^{-1}$ law is tested quantitatively in Sec.~\ref{subsec:spectralchar}. In the integrable limit $\delta_3\to0$ this channel closes and the bound state merely breathes (Sec.~\ref{subsec:fission}). Only the $\delta_3$ correction is retained, with Raman scattering and self-steepening explicitly omitted: this is the minimal integrability-breaking perturbation that isolates the resonant (Cherenkov) emission channel responsible for the entropy production analyzed here. A quantitative treatment of the full GNLSE with Raman and self-steepening, which superpose further irreversible channels (soliton self-frequency shift, optical shock) in the same direction, is the natural scope of follow-on work \cite{Agrawal2013,Husko2013,Younas2023}.

 The effective surface gravity $\kappa(\xi)$ is defined, in analogy with the acoustic-horizon construction, as the comoving-time gradient of the soliton-induced index profile at the horizon point,
\begin{equation}
\label{eq:surfgrav}
\kappa(\xi)=\frac{c}{2\,n_g^{\,2}}\left.\frac{\partial\,\Delta n(\tau,\xi)}{\partial\tau}\right|_{\tau=\tau_h(\xi)} .
\end{equation}
Each ingredient of Eq.~\eqref{eq:surfgrav} is fixed as follows. The Kerr index perturbation $\Delta n(\tau,\xi)=n_2 I(\tau,\xi)$ is the comoving refractive-index profile written by the instantaneous pulse intensity $I\propto|\psi|^2$; $n_g$ is the background group index of the fiber mode; and $\tau_h(\xi)$ is the horizon point, the comoving time at which the local group velocity of weak dispersive waves equals the soliton velocity. Equation~\eqref{eq:surfgrav} is then the direct optical counterpart of the acoustic surface gravity: it measures how steeply the effective flow profile crosses the horizon. Numerically, $\kappa(\xi)$ is obtained along the propagation by locating $\tau_h(\xi)$ from the group-velocity-matching condition and evaluating the gradient of $\Delta n\propto|\psi|^2$ at that point. Two clarifications delimit its role in this work. First, $\kappa(\xi)$ is a purely kinematic measure of the instantaneous horizon strength: it enters none of the entropy definitions of Sec.~\ref{sec:formalism} and plays no role in the GSL test. Second, whether the radiation accompanying a given $\kappa$ is thermal is a logically separate question, addressed in Sec.~\ref{subsec:spectralchar}.

Equation~\eqref{eq:dimensionless_gnlse} is integrated using the symmetrized Split-Step Fourier Method (SSFM), the standard numerical technique for solving nonlinear pulse propagation problems in dispersive media \cite{Agrawal2013}. To preserve the Hamiltonian structure of the dynamics and ensure photon number conservation during the rapid spectral broadening phase, we employ an adaptive step-size algorithm governed by a local error bound of $10^{-6}$ \cite{Sinkin2003}.

The system--bath separation required for the operational entropy measures is implemented via a dynamically centered super-Gaussian spectral filter $W_S(k)$ that tracks the instantaneous soliton peak, thereby defining the soliton-dominated domain $\Omega_S$ and its complementary radiation domain $\Omega_R$ without imposing artificial dissipation.

\subsection{Simulation Parameters}
\label{subsec:params}
We simulate a physically realizable regime characteristic of supercontinuum generation in solid-core photonic crystal fibers. Such systems allow for precise dispersion engineering, where the zero-dispersion wavelength (ZDW) and nonlinear coefficient can be tailored via core geometry to ensure robust soliton fission \cite{Dudley2006, Yuksel2025}. While the equations are solved in dimensionless form to isolate universal behavior, we present results in physical units for experimental clarity. The fiber parameters are chosen to match standard anomalous dispersion profiles: $\beta_2 = -15\,\mathrm{ps^2/km}$, $\beta_3 = 0.1\,\mathrm{ps^3/km}$, and $\gamma = 0.1\,\mathrm{W^{-1}km^{-1}}$. The input pulse duration is $T_0=50\,\mathrm{fs}$ with a soliton order of $N=3.5$, ensuring the system evolves deep into the fission regime where resonant radiation is efficiently generated \cite{Husko2013}.

\subsection{Soliton Breathing, Fission, and the Coarse-Graining Scale}
\label{subsec:fission}
The dynamical vocabulary used throughout---\emph{breathing}, \emph{fission}, and the coarse-graining scale $\Delta\xi_c$---is fixed as follows. The input pulse $\psi(0,\tau)=N\,\mathrm{sech}\,\tau$ launches a higher-order (bound-state) soliton of order $N$, defined in physical units by $N^2=\gamma P_0 T_0^2/|\beta_2|$ with $P_0$ the peak power \cite{Agrawal2013}. For $N>1$ and $\delta_3=0$ the evolution is exactly periodic: the bound state undergoes cyclic temporal compression and spectral broadening, recovering its initial shape after each soliton period $z_0=(\pi/2)\,L_D$, i.e., $\xi_0=\pi/2$ in the normalization of Eq.~\eqref{eq:dimensionless_gnlse}, where $L_D=T_0^2/|\beta_2|$ is the dispersion length \cite{Satsuma1974,Agrawal2013}. This reversible internal oscillation is what we call \emph{breathing}; it is the origin of the sign-indefinite excursions of $dS_{\mathrm{tot}}/d\xi$ analyzed in Sec.~\ref{subsec:gsl}.

Third-order dispersion breaks the velocity degeneracy that binds the constituents together \cite{Wai1986,Kodama1987}: near the first compression point, at $\xi_{\mathrm{fiss}}\simeq 1/N$ ($\approx0.29$ for $N=3.5$), the bound state splits into fundamental solitons---\emph{soliton fission} \cite{Herrmann2002,Dudley2006}. The constituents inherit the Zakharov--Shabat spectrum of the input pulse: the $j$-th ejected soliton has peak power $P_j/P_0=(2N-2j+1)^2/N^2$ and duration $T_j=T_0/(2N-2j+1)$ \cite{Satsuma1974,Dudley2006}, so that $N=3.5$ yields three constituents of relative amplitudes $6/3.5$, $4/3.5$, and $2/3.5$. The strongest, maximally compressed constituent dominates the resonant-radiation emission analyzed in Sec.~\ref{subsec:spectralchar}. All post-fission diagnostics in Fig.~\ref{fig:thermodynamics}(b) are accordingly evaluated on the window $\xi\ge2\gg\xi_{\mathrm{fiss}}$.

The coarse-graining scale entering Eq.~\eqref{eq:gsl_coarse} is determined from the simulation data rather than prescribed. We define $\Delta\xi_c$ as the dominant period of the oscillatory component of $S_{\mathrm{tot}}(\xi)$ in the post-fission window, extracted from the first zero crossing of the normalized autocorrelation of $dS_{\mathrm{tot}}/d\xi$ (equivalently, from the low-frequency peak of its power spectrum). For the reference run ($N=3.5$, $\delta_3=\DXICDELTA$) this yields $\Delta\xi_c=\DXICVAL$. The coarse-grained statements of Sec.~\ref{subsec:gsl_def} are insensitive to this choice: $\langle dS_{\mathrm{tot}}/d\xi\rangle_{\Delta\xi}\ge0$ for every window $\Delta\xi\ge\Delta\xi_c$, and $\eta_{\mathrm{GSL}}$ evaluated on the window-averaged rate varies by less than $\DXICROB\%$ as $\Delta\xi$ is varied over $[\Delta\xi_c,\,4\Delta\xi_c]$.

\section{Results and Discussion}
\label{sec:results}

By integrating the dimensionless GNLSE [Eq.~\eqref{eq:dimensionless_gnlse}] along the normalized propagation coordinate $\xi$, we obtain the full spectral evolution of the optical field $\tilde{\psi}(k,\xi)$. The operational entropy measures introduced in Sec.~\ref{sec:formalism} are then applied to this evolving spectrum in order to characterize the redistribution of information between subsystems and to assess the validity of the GSL under soliton fission dynamics.

\subsection{Spectral Partitioning and Entropy Generation}

\begin{figure}[htbp]
    \centering
    \IfFileExists{fig_evolution_optical_event_horizon.pdf}{
        \includegraphics[width=1.0\linewidth]{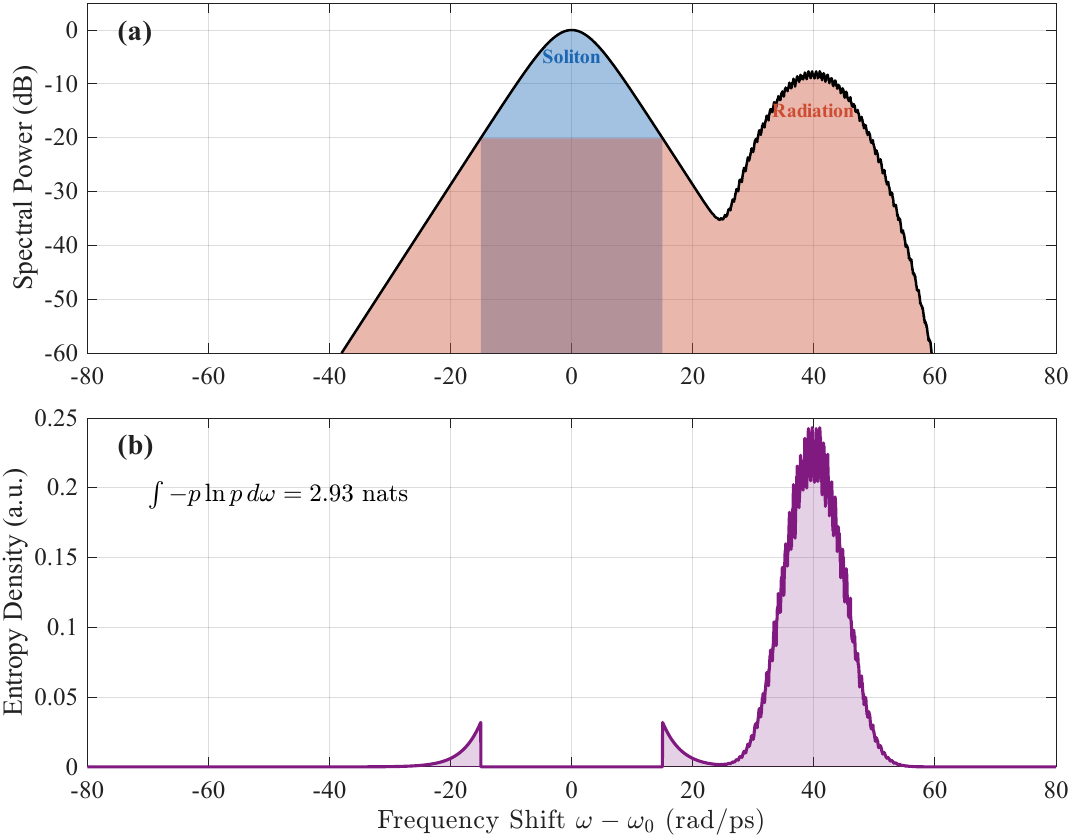}
    }{
        \framebox{\parbox{0.95\linewidth}{\centering
            \vspace{3cm}
            \textbf{[Image File: fig\_evolution\_optical\_event\_horizon.pdf]} \\
            \textit{Spectral Partitioning and Entropy Density} \\
            (Ensure file is in the same directory)
            \vspace{3cm}
        }}
    }
    \caption{\textbf{Spectral partitioning and entropy density of the optical horizon system.}
    (a) Bipartite decomposition of the instantaneous field spectrum into a localized soliton component (system) and a dispersive radiation component (bath). The shaded regions define the frequency domains $\Omega_S$ and $\Omega_R$ used in the operational entropy measures.
    (b) Spectral entropy density $-p(\omega)\ln p(\omega)$ evaluated over the radiation domain $\Omega_R$.}
    \label{fig:global_evolution}
\end{figure}

Figure~\ref{fig:global_evolution} presents a representative snapshot of the spectral and informational structure of the optical horizon system at a fixed propagation stage. Panel (a) shows the bipartite spectral partition of the total field into a narrowband soliton component, identified with the effective horizon subsystem, and a spectrally separated dispersive radiation bath. This physically motivated decomposition defines the disjoint integration domains $\Omega_S$ and $\Omega_R$ employed in the operational definitions of the horizon entropy $S_{\mathrm{hor}}$ and the radiation entropy $S_{\mathrm{rad}}$.

Panel (b) displays the corresponding spectral entropy density $-p(\omega)\ln p(\omega)$ evaluated over the radiation domain $\Omega_R$, where $p(\omega)$ is the normalized spectral intensity. The entropy density is strongly localized within the resonant radiation band, demonstrating that entropy production is dominated by horizon-induced dispersive emission rather than by the coherent soliton core. The finite integrated entropy,
\begin{equation}
\int_{\Omega_R} -p(\omega)\ln p(\omega)\, d\omega = 2.93 \ \text{nats},
\end{equation}
provides a quantitative measure of coarse-grained information redistribution arising from unresolved spectral microstructure in the radiation field. The numerical value depends on the fixed spectral resolution and bandwidth defining $\Omega_R$, consistent with an operational entropy.

 Figure~\ref{fig:global_evolution} is purely diagnostic: it characterizes the instantaneous subsystem partitioning and the associated entropy content at a given propagation coordinate. The figure does not, by itself, establish entropy monotonicity or the GSL. Rather, it furnishes the well-defined subsystem decomposition and operational entropy measures that are subsequently tracked as functions of the propagation distance $\xi$ to assess irreversible entropy production dynamically.

\FloatBarrier
\subsection{Entropy Evolution and the GSL}
\label{subsec:gsl}

The validity of the GSL is tested by monitoring the evolution of the coarse-grained entropies along the propagation coordinate. As the soliton undergoes temporal compression and fission, non-adiabatic mode conversion leads to the continuous emission of dispersive radiation. This process is accompanied by a dynamically varying dimensionless surface gravity $\kappa(\xi)$,  a kinematic measure of the instantaneous horizon strength.

\begin{figure}[htbp]
    \centering
    \IfFileExists{fig_sweep_robustness.pdf}{
        \includegraphics[width=\textwidth]{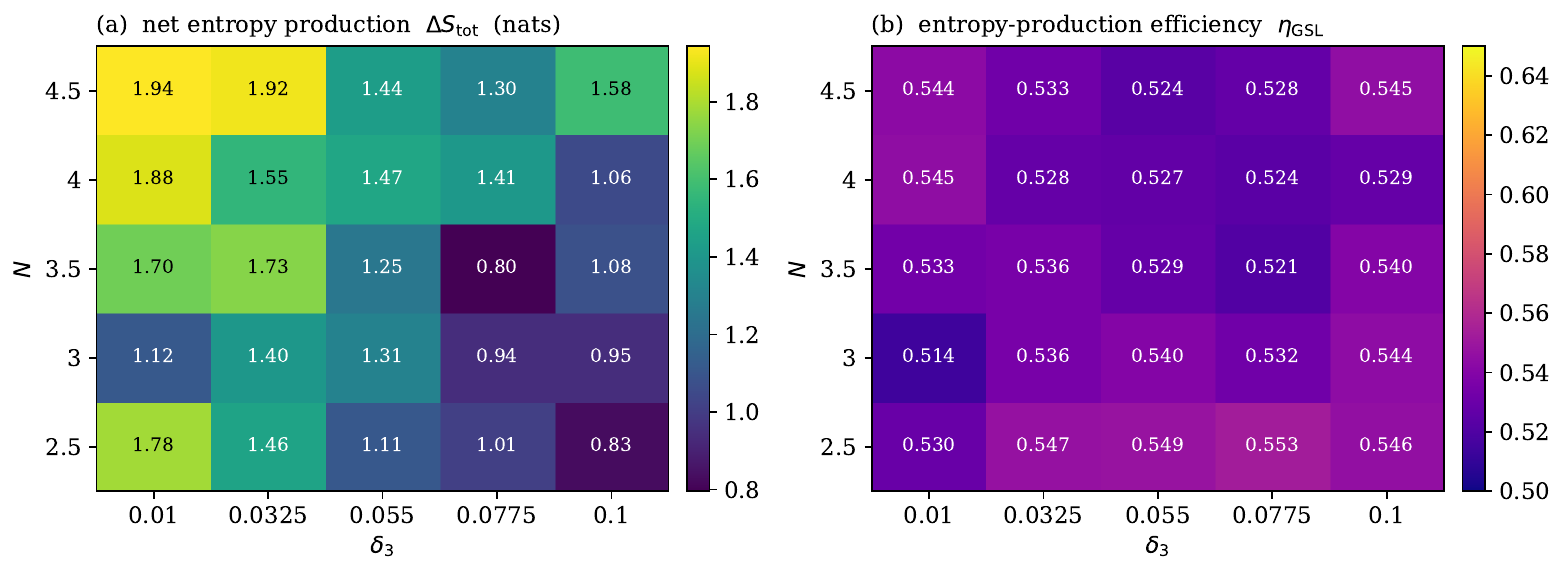}
    }{
        \framebox{\parbox{1.0\linewidth}{\centering
            \vspace{3cm}
            \textbf{[Image File: fig\_sweep\_robustness.pdf]} \\
            \textit{Robustness of the coarse-grained GSL across the simulation campaign} \\
            (Export from Supplementary Figs.~S4--S5)
            \vspace{3cm}
        }}
    }
    \caption{\textbf{Robustness of the coarse-grained GSL across the simulation campaign.}
    Net entropy production $\Delta S_{\mathrm{tot}}$ and entropy-production efficiency $\eta_{\mathrm{GSL}}$ over the $5\times5$ grid in $(N,\delta_3)$ specified in Sec.~\ref{subsec:gsl}, shown for the sech input and the comoving-envelope mask; the remaining pulse shapes and masking schemes yield the same qualitative picture (Supplementary Material, Sec.~S4). Every sampled configuration satisfies $\Delta S_{\mathrm{tot}}>0$ and $\eta_{\mathrm{GSL}}>\tfrac12$.}
    \label{fig:sweep}
\end{figure}

While the effective horizon subsystem  reorganizes its spectral content during radiation emission, the entropy generated in the radiation bath  dominates the balance. Tracking the total coarse-grained entropy,
\begin{equation}
S_{\mathrm{tot}}(\xi) = S_{\mathrm{hor}}(\xi) + S_{\mathrm{rad}}(\xi),
\end{equation}

we find that the trajectory is \emph{not} pointwise monotonic: roughly half of the local steps satisfy $dS_{\mathrm{tot}}/d\xi<0$, reflecting the reversible breathing of the soliton spectrum defined in Sec.~\ref{subsec:fission} [Fig.~\ref{fig:thermodynamics}(a,b)]. The coarse-grained law, however, is robustly satisfied across the full simulation campaign, which we now specify. The campaign comprises a $5\times5$ grid in the $(N,\delta_3)$ plane---five soliton orders $N\in\{\NGRIDVALS\}$ crossed with five dispersion strengths $\delta_3\in\{\DGRIDVALS\}$---each propagated for four input pulse shapes,
\begin{equation*}
\psi(0,\tau)\in\Bigl\{\,N\,\mathrm{sech}\,\tau,\;\; N e^{-\tau^2/2},\;\; N e^{-\tau^{8}/2},\;\; N\,\mathrm{sech}(\tau)\,e^{-iC\tau^2/2}\,\Bigr\}
\end{equation*}
(sech, Gaussian, super-Gaussian of order $m=4$, and chirped sech with chirp parameter $C=\CHIRPVAL$), and for three independent soliton/radiation masking schemes (static, centroid-tracking, and comoving-envelope), i.e., $5\times5\times4\times3=300$ configurations in total; the per-configuration protocol and results are reported in the Supplementary Material (Sec.~S4, Figs.~S2--S5) and summarized in Fig.~\ref{fig:sweep}. Every configuration yields a strictly positive net production,
\begin{equation}
\label{eq:sweep}
\Delta S_{\mathrm{tot}} = S_{\mathrm{tot}}(\xi_f)-S_{\mathrm{tot}}(0) > 0,
\qquad
\Delta S_{\mathrm{tot}} \in [\,0.82,\; 2.42\,]\ \text{nats},
\end{equation}
with an entropy-production efficiency $\eta_{\mathrm{GSL}}\approx 0.52\text{--}0.62$ everywhere, i.e., consistently above the unbiased-walk value $\tfrac{1}{2}$. The coarse-grained GSL therefore holds for every sampled configuration, not for one tuned trajectory. The optical event horizon thus behaves as a dissipative nonequilibrium structure consistent with the GSL expressed in terms of operational entropy.

 This result is robust with respect to soliton order, provided fission occurs.  The soliton therefore amplifies entropy, converting an initially coherent pump into spectrally delocalized radiation. This behavior mirrors the growth of entanglement entropy observed in acoustic horizon experiments, where entropy production arises from subsystem tracing rather than equilibration \cite{Steinhauer2016}.

\FloatBarrier

\subsection{ Spectral Spread versus Radiated Amplitude}
\label{subsec:spread}

The entropy measures and the resonant-radiation amplitude probe distinct properties of the same emission and should not be conflated. The classic Akhmediev--Karlsson analysis predicts that the \emph{amplitude} of solitonic Cherenkov radiation is exponentially suppressed with inverse dispersion strength, scaling as $\exp(-1/\delta_3)$ in the appropriate regime \cite{Akhmediev1995}. The radiation entropy $S_{\mathrm{rad}}$, by contrast, is the Shannon entropy of a \emph{normalized} spectral distribution and is therefore invariant under an overall rescaling of the radiated field: it measures the spectral \emph{spread} of the emission, not its power. Consequently $S_{\mathrm{rad}}$ does not, and should not, track the Akhmediev--Karlsson exponential. Empirically, $\Delta S_{\mathrm{tot}}$ varies by less than a factor of two across $\delta_3\in[0.01,0.10]$, whereas the Cherenkov amplitude factor varies over tens of orders of magnitude in the same interval, and with the opposite monotonicity. Once integrability breaking populates the dispersive continuum, the coarse-grained spectral redistribution rapidly saturates and becomes only weakly sensitive to $\delta_3$, while the radiated power remains sharply non-perturbative. The two quantities are complementary diagnostics of the same horizon process and must not be conflated: entropy production measures the breadth of the dispersive continuum, not the strength of the Cherenkov channel that seeds it. Semiclassical horizon analyses in which Hawking emission is controlled by barrier penetration probabilities \cite{Parikh2000} therefore remain a structural analogy for the present mechanism rather than a quantitative match.

\subsection{ Spectral Character of the Emitted Radiation}
\label{subsec:spectralchar}

\begin{figure}[htbp]
    \centering
    \IfFileExists{fig_thermo_analysis_sol.pdf}{
        \includegraphics[width=\textwidth]{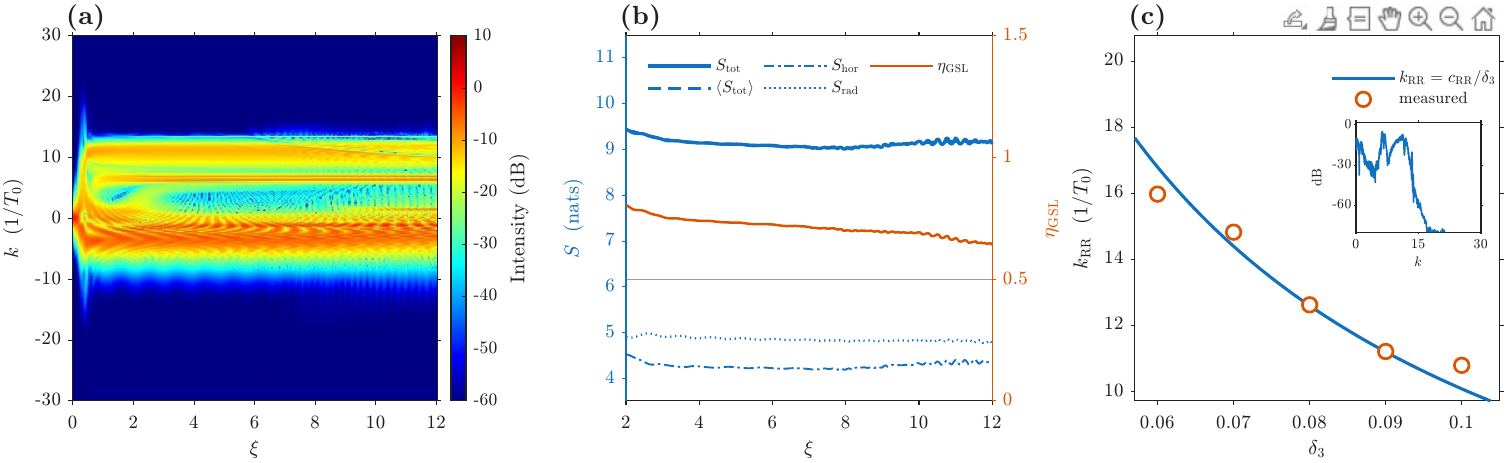}
    }{
        \framebox{\parbox{1.0\linewidth}{\centering
            \vspace{3cm}
            \textbf{[Image File: fig\_thermo\_analysis\_sol.pdf]} \\
            \textit{Thermodynamic Analysis of Soliton Fission} \\
            (Ensure file is in the same directory)
            \vspace{3cm}
        }}
    }
    \caption{\textbf{Thermodynamic signatures of a solitonic optical event horizon.}
    (a)  Spectral evolution $|\tilde{\psi}(k,\xi)|^2$ of the optical field along the propagation coordinate $\xi$; vertical axis is the spectral wavenumber $k$.
    (b)  Coarse-grained entropies $S_{\mathrm{hor}}$, $S_{\mathrm{rad}}$, total $S_{\mathrm{tot}}=S_{\mathrm{hor}}+S_{\mathrm{rad}}$, average $\langle S_{\mathrm{tot}}\rangle$ (left axis, nats), and efficiency $\eta_{\mathrm{GSL}}$ (right axis), over the post-fission window $\xi\ge2$.
    (c)  Measured resonant-radiation lobe position $k_{\mathrm{RR}}$ versus $\delta_3$ (symbols) and the scaling $k_{\mathrm{RR}}=c_{\mathrm{RR}}/\delta_3$ (line); inset, a representative spectrum $|\tilde{\psi}(k)|^2$ in dB.}
    \label{fig:thermodynamics}
\end{figure}

We now examine the spectral structure of the emitted radiation and, in particular, whether it admits a thermal (Boltzmann) interpretation with an associated effective temperature. This question must be posed carefully, because the radiation channel responsible for the entropy production analyzed above is the integrability-breaking resonant (Cherenkov) radiation, whose spectral form is governed by a phase-matching condition rather than by a thermal occupation. For the truncated GNLSE the resonant radiation is emitted at a wavenumber fixed by the soliton--radiation phase-matching condition, Eq.~\eqref{eq:phasematch}; for $\delta_3\ll1$ this predicts $k_{\mathrm{RR}}(\delta_3)\simeq c_{\mathrm{RR}}/\delta_3$ with an $\mathcal{O}(1)$ coefficient $c_{\mathrm{RR}}$ set by the cubic--quadratic balance and by the amplitude of the radiating sub-soliton in the fission regime \cite{Wai1986,Akhmediev1995,Skryabin2010}.
Direct extraction of the lobe position confirms this scaling [Fig.~\ref{fig:thermodynamics}(c)]: a through-origin fit over $\delta_3\in[0.06,0.10]$ gives $c_{\mathrm{RR}}=1.01$ ($R^2=0.93$), with the invariant $k_{\mathrm{RR}}\,\delta_3=1.02$ constant to within $5\%$ (a clean $\delta_3^{-1}$ law). This lobe is \emph{not} a decaying exponential and admits no stable, resolution-independent single-slope fit: this was verified with three independent extraction procedures (a fixed absolute-wavenumber tail fit, an automatically located lobe fit on the log-spectral envelope, and a fit in the soliton co-moving frequency frame) across $\delta_3\in[0.06,0.10]$ and a range of post-fission distances. The spectrum is a coherent phase-matched resonance, not a thermal tail---the physically expected behavior, since solitonic resonant radiation is a deterministic, phase-locked emission, whereas a genuinely thermal (Hawking-like) optical spectrum requires mode conversion across a group-velocity horizon \cite{Philbin2008,Belgiorno2010,Robertson2012}, a mechanism distinct from the integrability-breaking channel studied here. We therefore do not assign a quantitative analog Hawking temperature to this channel; $\kappa(\xi)$ remains well defined kinematically. Thermodynamic-uncertainty arguments bound the precision of any such effective temperature in open systems \cite{Sonnino2024b}. The coarse-grained GSL is independent of whether the emitted spectrum is thermal: spectral thermality and irreversibility are logically separate properties of this radiation channel. The thermal-spectrum question is well posed only for the group-velocity-horizon mechanism and is deferred to a dedicated treatment.

\FloatBarrier

\subsection{ Analog Backreaction and Soliton Recoil}
Finally, we note that the entropy production observed here is intrinsically linked to the dynamical backreaction of the radiation on the horizon. In our simulations, the emission of resonant radiation causes the soliton to lose energy and redshift (spectral recoil), altering its group velocity and thus the effective metric geometry. This mechanism is the optical equivalent of the ``quantum backreaction'' effects recently investigated in Bose--Einstein condensates, where vacuum fluctuations modify the background hydrodynamic flow \cite{Baak2022}. Our thermodynamic framework  indicates that this backreaction is thermodynamically consistent with the coarse-grained GSL: the horizon recoil balances the net entropy generated by the radiation, in keeping with the conservation laws governing the analog spacetime.

\section{Conclusion}
We have formulated a self-consistent nonequilibrium thermodynamic description of optical event horizons arising during soliton fission in nonlinear dispersive media. By defining operational, coarse-grained entropy measures for the coherent soliton subsystem (identified with the effective horizon) and for the incoherent dispersive radiation generated through integrability-breaking dynamics, we demonstrated that the total entropy obeys the GSL in a coarse-grained sense, even though the underlying field evolution is governed by a Hamiltonian equation of motion.  Nonlinear fiber-based optical horizons therefore behave as nonequilibrium thermodynamic systems rather than purely kinematic or geometric analogs.

 Conceptually, our results clarify that integrability breaking is an intrinsically irreversible, entropy-producing process, and that this irreversibility is logically independent of whether the emitted radiation is thermal: the integrability-breaking channel emits a coherent phase-matched resonance rather than a Hawking-like thermal spectrum, the latter being well posed only for the distinct group-velocity-horizon mechanism. Entropy production is shown to originate from coarse-graining over dynamically generated radiation degrees of freedom, providing a concrete physical mechanism for irreversibility that does not rely on dissipation or coupling to an external bath.  Soliton fission thus generates entropy intrinsically, in close parallel with semiclassical horizon physics, where entropy growth arises from subsystem tracing rather than equilibration.

 Experimentally, the entropy measures defined here are computable from a measured spectrogram alone; the coarse-grained GSL can therefore be tested in existing supercontinuum fiber setups with no extra diagnostics. Natural extensions of the present framework include the full GNLSE with Raman and self-steepening, non-soliton initial conditions such as modulation-instability or CW-breakup horizons, and the group-velocity-horizon configuration, for which a quantum thermal occupation is in principle well posed and which we treat separately.

\FloatBarrier
\section*{Data availability statement}
The simulation and analysis code, together with the data that support the findings of this study, are openly and persistently available at \doi{10.5281/zenodo.20713660}.

\section*{Acknowledgements}
The author gratefully acknowledges the support of the Material Simulation Laboratory at Pamukkale University. The computational resources and research environment provided by the laboratory were instrumental in the completion of this work. The author also thanks U.~R.~Fischer, H.~Zor~Oguz and Z.~M.~Yuksel for valuable discussions and insightful comments. The author is grateful to the anonymous referees for their insightful comments and constructive suggestions, which have significantly improved the quality of the manuscript. This work was supported by the Istanbul Okan University Scientific Research Projects Coordination Unit (IOU BAP) under Project No. OBAP2026010006, and by the Pamukkale University Scientific Research Projects Coordination Unit (PAU BAP) under Project No. 2025ALDEP037.

\bibliographystyle{unsrtnat}
\bibliography{references}

\end{document}